\documentclass[aps,prb,showpacs,twocolumn,letterpaper]{revtex4}

\usepackage{graphicx} 
\usepackage{amsfonts,amssymb,amsmath,amsthm}
\usepackage{bm}

\begin{document}
 
\title{Influence of strain and oxygen vacancies on the magnetoelectric
properties of multiferroic bismuth ferrite}

\date{\today}
 
\author{Claude Ederer} 
\affiliation{Materials Research Laboratory and Materials Department,
  University of California, Santa Barbara, CA 93106, U.S.A.}
\email{ederer@mrl.ucsb.edu}
\author{Nicola A.~Spaldin}
\affiliation{Materials Research Laboratory and Materials Department,
  University of California, Santa Barbara, CA 93106, U.S.A.}

\begin{abstract}
The dependencies on strain and oxygen vacancies of the ferroelectric
polarization and the weak ferromagnetic magnetization in the
multiferroic material bismuth ferrite, BiFeO$_3$, are investigated
using first principles density functional theory calculations. The
electric polarization is found to be rather independent of strain, in
striking contrast to most conventional perovskite ferroelectrics. It
is also not significantly affected by oxygen vacancies, or by the
combined presence of strain and oxygen vacancies. The magnetization is
also unaffected by strain, however the incorporation of oxygen
vacancies can alter the magnetization slightly, and also leads to the
formation of Fe$^{2+}$. These results are discussed in light of recent
experiments on epitaxial films of BiFeO$_3$ which reported a strong
thickness dependence of both magnetization and polarization.
\end{abstract}

\pacs{71.15.Mb, 75.70.Ak, 77.80.-e}
 
\maketitle

\section{Introduction}

Materials that simultaneously show electric and magnetic order are
currently gaining more and more attention. This is partly due to the
fact that such \emph{multiferroics} are promising materials for new
types of multifunctional device applications, but also because of the
interesting physics found in this class of materials. For example a
strong coupling between ferroelectric and antiferromagnetic domain
walls has been found in YMnO$_3$, \cite{Fiebig_et_al:2002} in
orthorhombic TbMnO$_3$ and TbMn$_2$O$_4$ the ferroelectric
polarization can be reoriented by a magnetic field,
\cite{Kimura_et_al_Nature:2003,Hur_et_al:2004} and ferromagnetic
ordering can be ``switched on'' by an electric field in hexagonal
HoMnO$_3$.  \cite{Lottermoser_et_al:2004}

Although magnetoelectric materials have been known for a long time,
\cite{Smolenskii/Chupis:1982} recent progress in thin-film growth and
other sample preparation techniques contributed considerably to their
renaissance. By using techniques such as pulsed laser deposition
(PLD), chemical vapor deposition (CVD), or molecular beam epitaxy
(MBE), many materials can nowadays be prepared as high quality
epitaxial thin films. One advantage of these techniques is the
possibility to stabilize otherwise metastable structures or to tune
material properties by varying the lattice mismatch between the film
and the substrate, thereby introducing epitaxial strain in the thin
film material.

Indeed, the effect of strain on the ferroelectric properties of
conventional ferroelectric materials is also a topic of current
interest. Strain effects can lead to a substantial increase of the
spontaneous polarization and Curie temperature, \cite{Choi_et_al:2004}
and even drive paraelectric materials (such as SrTiO$_3$) into the
ferroelectric phase. \cite{Haeni_et_al:2004} Since the mechanism for
ferroelectricity in multiferroic materials is often different from
that in conventional perovskite ferroelectrics,
\cite{Hill:2002,vanAken_et_al:2004} the question arises of whether
similar strain effects will be observed in multiferroics. Magnetic
properties can also be strongly affected by strain, mainly due to
large changes in anisotropy. \cite{Tsui_et_al:2000} Strain can also
affect the saturation magnetization and Curie
temperature. \cite{Gan_et_al:1998}

First principles density functional theory (DFT) calculations (see
e.g. Ref.~\onlinecite{Jones/Gunnarsson:1989}) play a crucial role in
studying the influence of strain on ferroelectric properties.
\cite{Neaton/Rabe:2003,Bungaro/Rabe:2004,Dieguez_et_al:2004} Epitaxial
strain can be introduced straightforwardly in DFT studies by fixing
the lattice parameters in the directions corresponding to the lateral
dimension of the substrate and allowing the system to relax in the
perpendicular direction. This makes it possible to clearly distinguish
between the effect of strain and other influences present in real
thin-film samples, such as interface effects or various types of
defects. Such information can then be used to optimize the properties
of the thin-film material.

In this work we study the influence of strain on the electric
polarization and magnetization of multiferroic bismuth ferrite,
BiFeO$_3$. Bismuth ferrite crystallizes in a rhombohedrally distorted
perovskite structure with space group $R3c$, \cite{Kubel/Schmid:1990}
where all ions are displaced along the [111] direction relative to the
ideal centrosymmetric positions, and the oxygen octahedra surrounding
the transition metal cations are rotated around this axis, alternately
clockwise and counterclockwise. The magnetic order is essentially
G-type antiferromagnetic \cite{Fischer_et_al:1980} but in addition,
the direction of the magnetic moments in the bulk rotates with a long
wavelength of 620~\AA
(Ref.~\onlinecite{Sosnowska/Peterlin-Neumaier/Streichele:1982}). We
have recently shown that if this spiral spin structure is suppressed,
the system shows \emph{weak ferromagnetism} \cite{Moriya:1960} with
the magnetic moments oriented perpendicular to the rhombohedral axis
and a slight canting of these magnetic moments resulting in a small
macroscopic magnetization. \cite{Ederer/Spaldin:2005}

Ferroelectric hysteresis loops have been measured for BiFeO$_3$ but
the experimental determination of an exact value for the spontaneous
polarization $P_s$ is difficult due to large leakage
currents. \cite{Teague/Gerson/James:1970} Several values are reported
in the literature, summarized in Table~I of
Ref.~\onlinecite{Neaton_et_al:2005}. A first principles calculation of
the spontaneous polarization of bulk BiFeO$_3$ results in a value of
$P_s \sim$ 95~$\mu$C/cm$^2$ (Ref.~\onlinecite{Neaton_et_al:2005});
recent measurements for epitaxial films grown on SrTiO$_3$ agree well
with this value. \cite{Wang_et_al:2003,Li_et_al:2004,Bai_et_al:2005}
These experiments also show a strong dependence of both magnetization
and ferroelectric polarization on the film
thickness. \cite{Wang_et_al:2003} One likely explanation for this
thickness dependence is the increase in strain with decreasing film
thickness; another is a change in the concentration of defects such as
oxygen vacancies. In this work we systematically examine both
possibilities using first principles DFT calculations.

Our main result is that the strain dependence of the ferroelectric
polarization in BiFeO$_3$ is rather weak compared with conventional
ferroelectric materials and that it cannot explain the variation of
the polarization reported for the thin film samples. The same is true
for the magnetization which also shows only weak strain
dependence. The weak strain dependence of the polarization is due to a
very stable ionic configuration in BiFeO$_3$, which manifests itself
in only small changes of the relative ionic positions when the lattice
is strained. 

The high stability of the ferroelectric configuration also leads to a
negligible dependence of the electric polarization on the oxygen
vacancy concentration. In contrast, we find that the magnetization of
BiFeO$_3$ is affected by the presence of oxygen vacancies but the
changes are not very systematic and depend on the precise position of
the oxygen vacancy. The presence of oxygen vacancies in all cases
leads to the formation of Fe$^{2+}$ ions, which can be identified
unequivocally in the partial densities of states although the actual
charge differences between the different Fe sites are very small.

\section{Method}
\label{sec:methods}

For this work we use first principles density functional theory (see
e.g. Ref.~\onlinecite{Jones/Gunnarsson:1989}) within the projector
augmented wave (PAW) method \cite{Bloechl:1994} as implemented in the
\emph{Vienna Ab-initio Simulation Package} (VASP)
\cite{Kresse/Furthmueller_PRB:1996,Kresse/Joubert:1999}. We include 15
valence electrons for Bi ($5d^{10}6s^26p^3$), 14 for Fe
($3p^63d^64s^2$), and 6 for each oxygen ($2s^22p^4$), use an energy
cutoff between 450-500~eV for the plane wave expansion of the PAWs, a
$4 \times 4 \times 4$ Monkhorst Pack grid of k-points,
\cite{Monkhorst/Pack:1976} and the tetrahedron method with Bl{\"o}chl
corrections for the Brillouin zone
integrations. \cite{Bloechl/Jepsen/Andersen:1994} For all structures
(unless otherwise noted) we relax the ionic positions while keeping
the lattice vectors fixed until the Hellman-Feynman forces are less
than 10$^{-2}$~eV/\AA. For the calculation of the local densities of
states at the Fe sites we use a sphere radius of 1.4~\AA. These values
have been found to give good convergence of all quantities under
consideration.

To treat exchange and correlation effects we use both the local spin
density approximation (LSDA) \cite{Jones/Gunnarsson:1989} and the
semiempirical LSDA+U method
\cite{Anisimov/Aryatesiawan/Liechtenstein:1997} for a better
description of the localized transition metal $d$ electrons. We have
recently shown that using the LSDA+U method and a moderate value of
$U$ = 3~eV (and $J$ = 1~eV) leads to a good description of the
structural parameters and the ferroelectric polarization in
BiFeO$_3$. \cite{Neaton_et_al:2005} Larger $U$ values shift the $d$
bands further down in energy relative to the oxygen $p$ states but
have only a small effect on the structural and ferroelectric
properties. $U$ = 3~eV can be regarded as a lower limit of what is
required to ensure the insulating character of BiFeO$_3$, and here we
exclusively use this value (and $J$ = 1~eV) in our LSDA+U
calculations. We do not claim that these values necessarily would also
lead to a good description of spectroscopic quantities such as
e.g. photoemission spectra.

There are two different LSDA+U approaches implemented in the VASP
code, (i) the traditional LSDA+U approach of Anisimov, Liechtenstein
and coworkers (in the so-called ``fully localized limit'')
\cite{Anisimov/Aryatesiawan/Liechtenstein:1997}, and (ii) a simplified
approach of Dudarev {\it et al.} \cite{Dudarev_et_al:1998} where only
the difference $U_\text{eff} = U - J$ enters. As shown in
Appendix~\ref{app:ldau} the latter approach (ii) is identical to
approach (i) when $J$ = 0, so that the difference between these two
approaches can be discussed in terms of a $J$-dependence. As pointed
out above, the structural and ferroelectric properties do not depend
strongly on the precise values of $U$ and $J$ and are therefore
basically identical for both approaches. In contrast, the
magnetization of BiFeO$_3$, which is due to a small canting of the
mainly antiferromagnetically coupled magnetic moments of the Fe
cations, \cite{Ederer/Spaldin:2005} is strongly
$J$-dependent. Although the absolute value of this canting (and
therefore the macroscopic magnetization) depends strongly on $J$, the
effects of strain and oxygen vacancies are basically independent of
the actual value of $J$. Since the focus of the present paper is on
these effects, we postpone the detailed analysis of the $J$ dependence
of the magnetization to a future publication and always present two
data sets for the magnetization, one obtained using $U$ = 3~eV and $J$
= 1~eV, the second obtained using $U$ = 2~eV and $J$ = 0~eV (or
equivalently approach (ii) with $U_\text{eff}$ = 2~eV). Our
conclusions regarding the effects of strain and oxygen vacancies on
the magnetization in BiFeO$_3$ apply independently to both data
sets. We point out that a significant $J$ dependence is only observed
for the \emph{canting} of the local magnetic moments of the Fe
cations. The absolute values of these magnetic moments are rather
independent of $J$, as long as $J$ is varied within reasonable limits
($J < 1.5$~eV).

For the calculation of the ferroelectric polarization we use the
Berry-phase approach developed by Vanderbilt and King-Smith.
\cite{King-Smith/Vanderbilt:1993,Vanderbilt/King-Smith:1994,Resta:1994}
In this theory the polarization of a periodic solid is represented by
a three-dimensional lattice, and experimentally accessible
polarization differences are obtained by connecting two points of the
``polarization lattices'' of the initial and final states, which can
be transformed into each other through a continuous ``path'' of
insulating intermediate states. In the present case special care must
be taken in determining which points of the initial and final state
polarization lattices have to be connected. First, because the value
of the polarization difference in BiFeO$_3$ is comparable to the
distance between neighboring points on the polarization lattices,
\cite{Neaton_et_al:2005} and in addition, because the direction of the
spontaneous polarization in the monoclinically strained structures and
in the supercells containing oxygen vacancies is not symmetry
restricted. In such cases the noncentrosymmetric distortions of the
ionic positions in the corresponding systems are gradually reduced
until an unambiguous connection can be made, i.e. the polarization of
some intermediate states is explicitly calculated.

\section{Results and Discussion}

\subsection{Strain dependence of the electric polarization for (111)
  oriented films}

\begin{table*}
\caption{Strain $\epsilon$, in-plane lattice parameter $a_\text{hex}$,
relaxed out-of-plane lattice parameter $c^{(0)}_\text{hex}$, volume
$V$ of the rhombohedral unit cell (containing two formula units),
absolute and relative displacements of Fe and O ions along the [111]
direction (compared to an ideal centrosymmetric reference structure
with the same lattice parameters, $u_i(\epsilon) = R_i(\epsilon) -
R_{i,0}(\epsilon)$, $\Delta u_i(\epsilon) =
(u_i(\epsilon)/u_i(0))-1$), spontaneous polarization $P_s$, and
magnetization $M_s$ for BiFeO$_3$ strained within the (111)
plane. Upper values are obtained using the LSDA, lower values are
obtained using the LSDA+U method. For the LSDA+U magnetization, the
first value refers to $U = 2$~eV, $J = 0$~eV and the second value
refers to $U = 3$~eV, $J = 1$~eV (see Sec.~\ref{sec:methods} for
details).}
\label{tab:111-strain}
\begin{center}
\begin{ruledtabular}
\begin{tabular}{c|ccc|cc|cc|c|c}
$\epsilon$ [\%] & 
$a_\text{hex}$ [\AA] & $c^{(0)}_\text{hex}$ [\AA] & $V$ [\AA$^3$] & 
$u_\text{Fe}$ [\AA] & $\Delta u_\text{Fe}$ [\%] & 
$u_\text{O}$ [\AA] & $\Delta u_\text{O}$ [\%] & 
$P_s$ [$\mu$C/cm$^2$] & $M_s$ [$\mu_\text{B}$/Fe] \\
\hline 
$-$3 & 5.326 & 13.68 & 112.01 & 0.294 & 16.8 & 0.527 & 2.5 & 102.8 & 0.04 \\ 
$-$1 & 5.431 & 13.45 & 114.53 & 0.268 &  6.2 & 0.520 & 1.2 & 100.1 & 0.05 \\
0    & 5.485 & 13.31 & 115.62 & 0.252 &  0   & 0.514 & 0   &  98.9 & 0.05 \\
+1   & 5.541 & 13.22 & 117.16 & 0.246 & -2.5 & 0.517 & 0.6 &  98.4 & 0.05 \\
+3   & 5.655 & 13.00 & 120.01 & 0.227 & -9.7 & 0.520 & 1.3 &  97.9 & 0.05 \\
\hline
$-$3 & 5.343 & 13.92 & 114.71 & 0.357 & 21.8 & 0.573 & 6.2 &  97.7 &
0.02/0.11 \\
0    & 5.503 & 13.48 & 117.86 & 0.293 &  0   & 0.539 & 0   &  94.0 &
0.03/0.10
\end{tabular}
\end{ruledtabular}
\end{center}
\end{table*}

We first investigate the effect of strain corresponding to a (111)
orientation of the substrate. This geometry preserves the rhombohedral
symmetry found in the unstrained system and the spontaneous
polarization remains oriented along the [111] direction. We consider
compressive strain as well as tensile strain. We fix the nearest
neighbor distance $a_\text{hex}$ between identical cations within the
(111) plane to the values shown in Table~\ref{tab:111-strain} and then
vary the out-of-plane lattice parameter $c_\text{hex}$ while relaxing
all ionic positions.\footnote{The rhombohedral lattice can be
described either using a rhombohedral or a hexagonal setup (see
e.g. Ref.~\onlinecite{ITables}). We use one or the other notation
depending on what is appropriate in the given context.} The
spontaneous electric polarization $P_s$ and magnetization per Fe
cation $M_s$ are calculated for the relaxed value of the out-of-plane
parameter, $c^{(0)}_\text{hex}$, corresponding to the minimum of the
total energy with fixed in-plane parameter $a_\text{hex}$ (and relaxed
ionic coordinates). The results are summarized in
Table~\ref{tab:111-strain}. The strain is defined as $\epsilon =
a_\text{hex}/a_{\text{hex},0} - 1$ where $a_{\text{hex},0}$ is the
corresponding lattice constant for the unstrained system. For
comparison we mention that the lattice constant of SrTiO$_3$, which is
a commonly used substrate material, would lead to a compressive strain
of $\epsilon \approx -2$~\%.

One can see that compressive epitaxial strain in the (111) plane leads
to a reduction of the unit cell volume whereas tensile strain leads to
a volume increase, i.e. the system does not behave like an ideal
elastic medium. This result is similar to that found in
Refs.~\onlinecite{Neaton/Hsueh/Rabe:2002} and
\onlinecite{Neaton/Rabe:2003} for BaTiO$_3$. The spontaneous
polarization $P_s$ of BiFeO$_3$ increases slightly with increasing
compressive strain (4~\% increase of the polarization for $\epsilon =
-3$~\%) but the effect is much smaller than in the conventional
ferroelectric systems BaTiO$_3$ and PbTiO$_3$. In BaTiO$_3$ a strain
of $\epsilon = -1$~\% leads to an increase in the polarization of
$\sim$ 35~\% (Ref.~\onlinecite{Neaton/Hsueh/Rabe:2002}) whereas a
similar strain in PbTiO$_3$ increases the polarization by 15-20~\%.
\cite{Bungaro/Rabe:2004} The largest strain effect on the polarization
is seen in SrTiO$_3$ which is not ferroelectric in the unstrained
state but develops a spontaneous polarization in epitaxially strained
films. \cite{Haeni_et_al:2004}

\begin{figure}
\centerline{\includegraphics*[width=0.4\textwidth]
{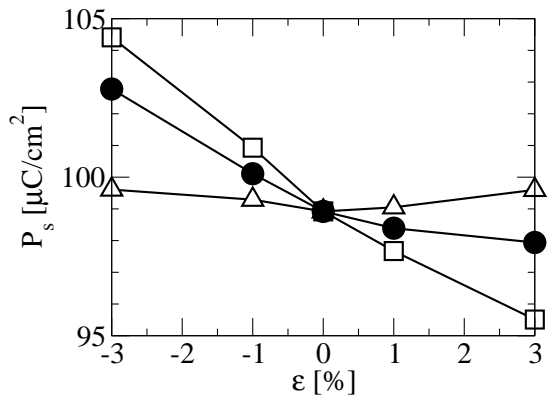}}
\caption{Polarization $P_s$ as function of strain $\epsilon$
  calculated within the Berry phase approach (full circles) and by a
  simple point charge model (see text) using the formal charges (open
  triangles) and Born-effective charges of the unstrained system (open
  squares).}
\label{fig:111-strain}
\end{figure}

We point out that the observed small change in spontaneous
polarization is consistent with the small changes in ionic
displacements that result from the applied strain. If we calculate the
expected change in polarization based on a simple point charge model
as $P_s(\epsilon) = P_s(\epsilon = 0) + \frac{1}{V(\epsilon)} \sum_i
Z_i (R_i(\epsilon) - R_i(\epsilon=0))$ (here $Z_i$ is the charge
associated with ion $i$ (see below), $R_i(\epsilon)$ is the
corresponding strain-dependent ionic position, $V$ is the unit cell
volume (also strain dependent), and the sum extends over all ions in
the unit cell) then the resulting change in polarization for
reasonable values of $Z_i$ is comparable to the change found by the
direct calculation of $P_s$ using the Berry-phase
approach. Fig.~\ref{fig:111-strain} shows the change in polarization
as a function of strain, calculated with the Berry-phase approach as
well as with the above formula using for $Z_i$ (i) the formal charges
and (ii) the Born effective charges (BECs) of the unstrained
structure. \cite{Neaton_et_al:2005} The weaker strain dependence of
the spontaneous polarization in BiFeO$_3$ compared to other
ferroelectrics can therefore be traced back to the weaker strain
dependence of the ionic displacements in this material.

Table~\ref{tab:111-strain} also shows the change in ionic
displacements (compared to an ideal centrosymmetric reference
structure with the same lattice parameters) as a function of
strain. The positions of the Bi cations are used as reference and
therefore the displacements of these ions are zero by definition. For
a compressive strain of $\epsilon = -3$~\% the displacements of the Fe
cations change only by $\sim$ 17~\% (LSDA) whereas the displacements
of the oxygen atoms are nearly strain-independent. For comparison, an
epitaxial strain of -2.28~\% in the (001) plane of tetragonal
BaTiO$_3$ leads to a change of the displacements of the Ti cations
along [001] of $\sim$ 52~\%, and the displacements of the O anions in
this case change by even more than
100~\%. \cite{Neaton/Hsueh/Rabe:2002}

There are three likely explanations for the weak strain dependence of
the ionic displacements in BiFeO$_3$. The first is the general high
stability of ferroelectrics with high Curie temperatures, large ionic
displacements, and large energy differences between the ground state
and the centrosymmetric reference structure. BiFeO$_3$ has a very high
Curie temperature ($T_C$ = 1123~K) and the energy gain in the ground
state $R3c$ structure is $\sim$ 0.25~eV per formula unit compared to
the centrosymmetric $R\bar{3}c$ structure and $\sim$ 1~eV compared to
the cubic perovskite structure). This explanation is consistent with
the fact that the strain dependence of the polarization is already
weaker in PbTiO$_3$ (Ref.~\onlinecite{Bungaro/Rabe:2004}, $T_C =
763$~K) than in BaTiO$_3$ (Ref.~\onlinecite{Neaton/Rabe:2003}, $T_C =
400$~K). If this explanation is valid, a similar strain independence
of the electric polarization should be observable for LiNbO$_3$ which
is isostructural to BiFeO$_3$ and has a even higher Curie temperature
of 1480~K (all Curie temperatures are taken from
Ref.~\onlinecite{LB:36}). The second possible explanation is that the
mechanism driving the ferroelectric distortion in BiFeO$_3$, namely
the stereochemically active lone pair, \cite{Seshadri/Hill:2001} is
relatively inert to the changes in the lattice vectors caused by
epitaxial strain. In conventional ferroelectric perovskites the
ferroelectric distortion is stabilized by charge transfer from the
oxygen into the unoccupied transition metal $d$
orbitals. \cite{Cohen:2000} This charge transfer mechanism is probably
more sensitive to small changes in bond lengths than the
stereochemical activity of the Bi lone electron pair. In this case the
study of the strain dependence of the electric polarization in the
multiferroic BiMnO$_3$ would be of interest. The ferroelectricity in
BiMnO$_3$ is driven by the Bi lone pair \cite{Seshadri/Hill:2001} but
its ``general stability'' is not large (reported $T_c \sim$~760~K,
Ref.~\onlinecite{Kimura_et_al_PRB:2003}) Thus a small (large) strain
dependence in LiNbO$_3$ and a large (small) strain dependence in
BiMnO$_3$ would confirm the ``general stability'' (lone pair) origin
of the polarization stability. A third possible explanation for the
weak strain dependence in BiFeO$_3$ is the special geometry of the
oxygen octahedra in the $R3c$ structure. In ferroelectrics like
BaTiO$_3$ and PbTiO$_3$ all ions are displaced only along the polar
direction whereas in BiFeO$_3$ (and also in LiNbO$_3$) the oxygen
octahedra are also rotated around this axis. The resulting geometry of
the oxygen cages surrounding the transition metal cations could be
less favorable for an additional strain-induced displacement of the
ions compared to the simpler geometry found in BaTiO$_3$ and
PbTiO$_3$. Future studies will shed more light on this issue.

The results obtained using the LSDA+U method are very similar to the
LSDA results. Although the relative change in displacements seems to
be slightly larger than within the LSDA (see
Table~\ref{tab:111-strain}), the relative changes in polarization are
exactly the same in both cases. This reflects the fact, which has been
already pointed out in Ref.~\onlinecite{Neaton_et_al:2005}, that the
explicit treatment of electronic correlations within the LSDA+U method
has only minor effect on the structural properties of this
system. Apart from resulting in a slightly larger equilibrium volume,
in better agreement with experimental data, the main effect is an
improved description of the electronic structure resulting in a larger
band gap and a stable insulating phase.

\subsection{Strain dependence of the polarization for (001) oriented films}
\label{monocl}

\begin{table}
\caption{Experimental lattice parameters (in \AA) found in
  representative BiFeO$_3$ films \cite{Ramesh:private} (line
  ``Exp.''), together with the corresponding lattice parameter
  $a_\text{rh}$ of rhombohedral bulk BiFeO$_3$
  (Ref.~\onlinecite{Kubel/Schmid:1990}), and the values used in the
  calculations for the monoclinically strained films (line
  ``Theo.''). $a$ and $b$ correspond to the two in-plane directions
  and $c$ to the out-of-plane direction of the monoclinic lattice. The
  values in the given form correspond to the lengths of the cube edges
  of the distorted cubic perovskite structure. Strain values
  $\epsilon$ are also given.}
\label{tab:monoclinic}
\begin{center}
\begin{ruledtabular}
\begin{tabular}{c|c|ccc|ccc}
& & \multicolumn{3}{c|}{``thin films''} & \multicolumn{3}{c}{``thick films''}
  \\ \cline{3-8} 
& $a_\text{rh}/\sqrt{2}$ & $a/\sqrt{2}$ & $b/\sqrt{2}$ & $c$ &
  $a/\sqrt{2}$ & $b/\sqrt{2}$ & $c$ \\ 
\hline 
Exp.  & 3.98 & 3.92 & 3.92 & 4.06 & 3.91 & 3.97 & 4.00 \\ 
Theo. & 3.89 & 3.83 & 3.83 & 3.97 & 3.82 & 3.88 & 3.91 \\ 
\hline 
\multicolumn{2}{c|}{$\epsilon$} & $-$1.5\% & $-$1.5\% & +2\% & $-$2\% &
  $-$0.2\% & +0.4\% 
\end{tabular}
\end{ruledtabular}
\end{center}
\end{table}

We now discuss the case of BiFeO$_3$ deposited on a (001) surface. In
this case the epitaxial constraint of the cubic substrate enforces
90$^\circ$ angles of the lattice vectors within the (001) planes, in
conflict with the rhombohedral distortion of bulk BiFeO$_3$. One can
expect that the competition between these two effects leads to a
base-centered monoclinic structure and indeed monoclinic symmetry is
found experimentally in epitaxial BiFeO$_3$ films deposited on a (001)
surface of SrTiO$_3$. \cite{Ramesh:private} The low symmetry of the
monoclinic structure renders a systematic computational investigation
unfeasible for this type of epitaxial strain. We have therefore
performed calculations for two structures, one representative of very
thin films ($\sim$ 100-200~nm), the other representative of thicker
films ($\sim$ 400~nm), based on experimentally determined lattice
parameters. \cite{Ramesh:private} To account for the fact that the
theoretically determined lattice parameters are usually slightly
different from the experimental lattice parameters we have scaled all
values accordingly, so that the strain in the calculations is the same
as that found experimentally. We use a monoclinic angle $\beta$ =
89.5$^\circ$, consistent with the experimental
data. \cite{Ramesh:private} All lattice parameters are summarized in
Table~\ref{tab:monoclinic}. As reference for the theoretical lattice
parameters we use the values obtained by relaxing the bulk
rhombohedral unit cell using
$U_\text{eff}$=2~eV. \cite{Neaton_et_al:2005} In the thin films the
large in-plane stress leads to a large out-of-plane relaxation and a
$\sqrt{2} c/a$ ratio that deviates significantly from 1. In the thick
films the in-plane stress is partially released because of the two
different in-plane lattice parameters $a$ and $b$, leading to
$\sqrt{2} c/a$ ratio (or $\sqrt{2} c/b$ ratio) closer to 1 and nearly
no strain in the out-of-plane direction. The results for the electric
polarization and magnetization calculated after relaxing all the ionic
positions are shown in Table~\ref{tab:mc-results}. To ensure the
insulating character of the systems we use the LSDA+U method for these
calculations.

\begin{table}
\caption{Calculated absolute values of the polarization $|P_s|$ and
  out-of-plane component $(P_s)_{[001]}$ (in $\mu$C/cm$^2$) for the
  monoclinic structures. $P_\text{exp}$ is the value measured in the
  [001] direction, \cite{Wang_et_al:2003} and should be compared to
  the calculated $(P_s)_{[001]}$. For the magnetization $M_s$, the
  first value refers to $U = 2$~eV, $J = 0$~eV and the second value
  refers to $U = 3$~eV, $J = 1$~eV (see Sec.~\ref{sec:methods} for
  details).}
\label{tab:mc-results}
\begin{center}
\begin{ruledtabular}
\begin{tabular}{r|cc|c|c}
& $|P_s|$ & $(P_s)_{[001]}$ & $P_\text{exp}$
(Ref.~\onlinecite{Wang_et_al:2003}) & $M_s$ [$\mu_\text{B}$/Fe] \\ 
\hline 
``thin films''  & 94.8 & 63.4 & 50-60 & 0.03/0.10 \\ 
``thick films'' & 92.1 & 57.0 & 25-30 & 0.03/0.11
\end{tabular}
\end{ruledtabular}
\end{center}
\end{table}

Due to the larger $c/a$ ratio in the ``thin film'' structure the
direction of the polarization rotates further away from the [111]
direction (towards the [001] direction) compared to the ``thick film''
structure. This leads to an increase in the out-of-plane component of
the polarization, in spite of the fact that the effect of strain on
the absolute value of the electric polarization is rather small. Thus,
experiments that measure the out-of-plane component of the
polarization can expect to see changes in $P$ due to this
re-orientation effect, even in cases where the strain-dependence of
the magnitude of the polarization is small.

Comparing the calculated values of the spontaneous polarization with
the experimental data for (001) oriented films of different thickness
from Ref.~\onlinecite{Wang_et_al:2003} (see
Table~\ref{tab:mc-results}) shows that the theoretical values agree
reasonably well with the experimental data for the thin films
(perpendicular component of the polarization). The reported
experimental value for the thicker films seems to be smaller than the
calculated value. This could be due to incomplete switching of the
polarization in the thicker films. 

In summary, the strong dependency of the polarization from the film
thickness reported in Ref.~\onlinecite{Wang_et_al:2003} is probably a
sum of two effects: (i) the rotation of the polarization away from the
[111] direction in the thinner films due to the increased $c/a$ ratio,
and (ii) incomplete switching of the polarization in the thicker
films.

\subsection{Strain dependence of the magnetization}

It can be seen from Tables~\ref{tab:111-strain} and
\ref{tab:mc-results} that there is no significant strain dependence of
the magnetization in BiFeO$_3$. The absolute value of the
magnetization (resulting from the canting of the Fe magnetic moments)
is significantly larger for the traditional $J$ dependent LSDA+U
treatment compared to the simplified approach corresponding to $J$ =
0~eV (see Sec.~\ref{sec:methods}) but in both cases this value is not
significantly changed by strain, either for the rhombohedral symmetry
(Table~\ref{tab:111-strain}) or for the monoclinically strained
structure (Table~\ref{tab:mc-results}). 

\subsection{Influence of oxygen vacancies on the electric and magnetic
  properties}

To investigate the influence of oxygen vacancies on the ferroelectric
and magnetic properties of BiFeO$_3$ we perform calculations for a
unit cell doubled along one of the rhombohedral lattice vectors (so
that it contains four formula units), in which we remove one oxygen
atom and then relax all ionic positions in the supercell. We then
calculate total and partial densities of states, electric
polarization, and magnetization. We do this for the rhombohedral bulk
structure as well as for the ``thin film'' monoclinically distorted
structure described in Section~\ref{monocl} (see
Table~\ref{tab:monoclinic}). The use of a doubled unit cell
artificially reduces the symmetry of the rhombohedral system to the
monoclinic space group $Bb$ (in the case of the monoclinically
distorted structure this is already the space group of the original
unit cell). In this lowered symmetry there are three inequivalent
groups of oxygen anions. Removing an oxygen anion from one of the
three groups in turn leads to three different arrangements of oxygen
vacancies. In the following we refer to these different configurations
(in a somewhat arbitrary way) as configurations I, II, and III. The
corresponding supercells have a vacancy concentration of 8.3~\%, which
means that one out of 12 oxygen anions is missing. This can formally
be written as BiFeO$_{3-\delta}$ with $\delta=0.25$.

\begin{table}
\caption{Polarization $|P_s|$ and magnetization $M_s$ calculated
  within the LSDA+U method for the systems with oxygen vacancies. The
  upper (middle) three lines correspond to the three vacancy
  configurations based on the rhombohedral bulk structure (monoclinic
  ``thin film'' structure). The last line corresponds to the tripled
  unit cell. For the magnetization, the first value refers to $U =
  2$~eV, $J = 0$~eV and the second value refers to $U = 3$~eV, $J =
  1$~eV (see Sec.~\ref{sec:methods} for details).}
\label{tab:O-vac}
\begin{center}
\begin{ruledtabular}
\begin{tabular}{c|ccc}
 & $|P_s|$ [$\mu$C/cm$^2$] & $M_s$ [$\mu_\text{B}$/Fe] \\ 
\hline 
I   & 96.6 & 0.01/0.07 \\ 
II  & 95.0 & 0.07/0.14 \\
III & 97.1 & 0.05/0.11 \\ 
\hline 
I   & 97.1 & 0.02/0.08 \\ 
II  & 94.6 & 0.07/0.14 \\ 
III & 99.9 & 0.05/0.10 \\ 
\hline 
tripled & 96.5 & 0.06/0.12
\end{tabular}
\end{ruledtabular}
\end{center}
\end{table}

All systems are metallic within the LSDA but become insulating within
the LSDA+U. Table~\ref{tab:O-vac} shows the values for the electric
polarization and magnetization for all supercells calculated within
the LSDA+U method. Remarkably, the electric polarization is not
significantly affected by the presence of oxygen vacancies and the
accompanying structural and electronic distortions. This again
reflects the high stability of the ferroelectric configuration in this
system. Even the combined presence of epitaxial strain and oxygen
vacancies in the monoclinically distorted structures does not lead to
significant changes in the ferroelectric polarization $P_s$.

\begin{figure}
\includegraphics*[width=0.4\textwidth]{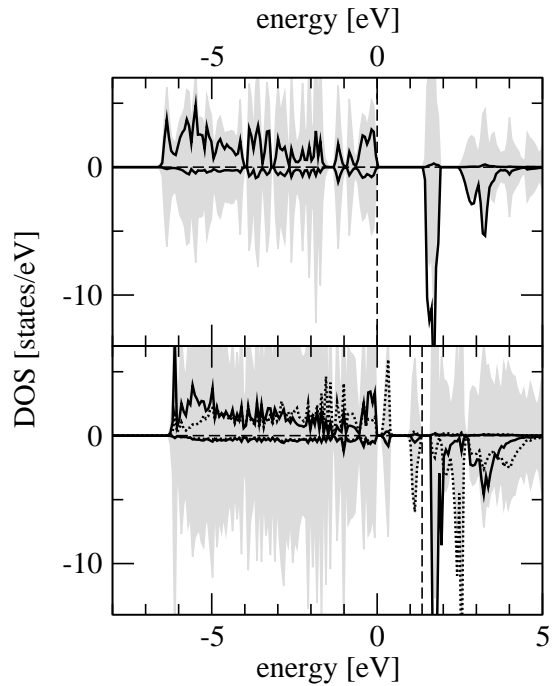}
\caption{Total (gray shaded) and partial Fe $d$ densities of states
 for the unstrained rhombohedral structure (upper panel) and for
 vacancy configuration I (lower panel) calculated for $U_\text{eff}$ =
 2~eV. Minority spin states are shown with a negative sign. The full
 lines correspond to the ``Fe$^{3+}$'' ions, the dotted lines in the
 lower panel corresponds to the ``Fe$^{2+}$'' ions. The dashed
 vertical lines indicate the highest occupied energy levels. Zero
 energy is set to the upper edge of the ``Fe$^{3+}$'' majority spin
 bands.}
\label{fig:dos}
\end{figure}

Figure~\ref{fig:dos} shows the densities of states (total and partial
Fe $d$) for vacancy configuration I based on the rhombohedral bulk
structure together with the corresponding densities of states for the
same structure without oxygen vacancies. The densities of states for
the other systems all look very similar to that shown in
Figure~\ref{fig:dos}. The four Fe cations in the supercell containing
the vacancy can be divided in two classes which we call ``Fe$^{3+}$''
and ``Fe$^{2+}$''. The partial $d$ densities of states for the two
``Fe$^{3+}$'' cations are very similar to the case without the oxygen
vacancies: the (local) majority spin states are completely filled and
the (local) minority spin states are completely empty (apart from a
small contribution arising from the hybridization with the O 2$p$
states), indicating a $d^5$ high-spin electronic configuration. The
unoccupied minority $d$ states are split into the $t_{2g}$ and $e_g$
manifolds characteristic of the predominantly octahedral symmetry of
the crystal field. The partial densities of states for the
``Fe$^{2+}$'' cations are significantly different from this. Here, the
$t_{2g}$ minority states are partially filled by approximately one
electron (indicating a high spin $d^6$ electron configuration) and
there is a small gap between the occupied and unoccupied minority
$t_{2g}$ states. The densities of states therefore suggest a picture
of distinct ``Fe$^{2+}$'' and ``Fe$^{3+}$'' cations with $d^5$ and
$d^6$ electron configurations respectively.

Although the presence of both ``Fe$^{2+}$'' and ``Fe$^{3+}$'' seems
rather obvious from the analysis of the densities of states, the local
charges, obtained by integrating the partial densities of states up to
the Fermi-energy, differ only slightly ($\sim 0.1~e$) for the two
types of Fe cations. Furthermore, these charges do not represent the
formal charges corresponding to the electron configurations mentioned
in the previous paragraph. The problem of assigning local (static)
charges based on the continuous electron density in periodic solids is
well known (see
e.g. Ref.~\onlinecite{Ghosez/Michenaud/Gonze:1998}). In practice,
local quantities are found by defining spheres around the atomic sites
and projecting the Bloch-functions onto a local basis. Since the radii
of such spheres are arbitrary within reasonable limits, and the
intermediate region between these spheres is either neglected or
double-counted, the local quantities defined in this way are not
uniquely defined. Furthermore, due to the formation of band-states
within a periodic solid, the angular momentum $l$ is no longer a good
quantum number. Even a pure ``$d$ band'', constructed as the Bloch-sum
of localized $d$ orbitals, usually contains some $s$ and $p$ character
when projected onto a local $l$ basis. It is therefore clear that
locally defined charges are poor indicators for the oxidation states
of the ions. The clear qualitative differences in the local densities
of states appear to be much more appropriate for this
purpose.

\begin{figure}
\centerline{
\includegraphics*[width=0.20\textwidth]{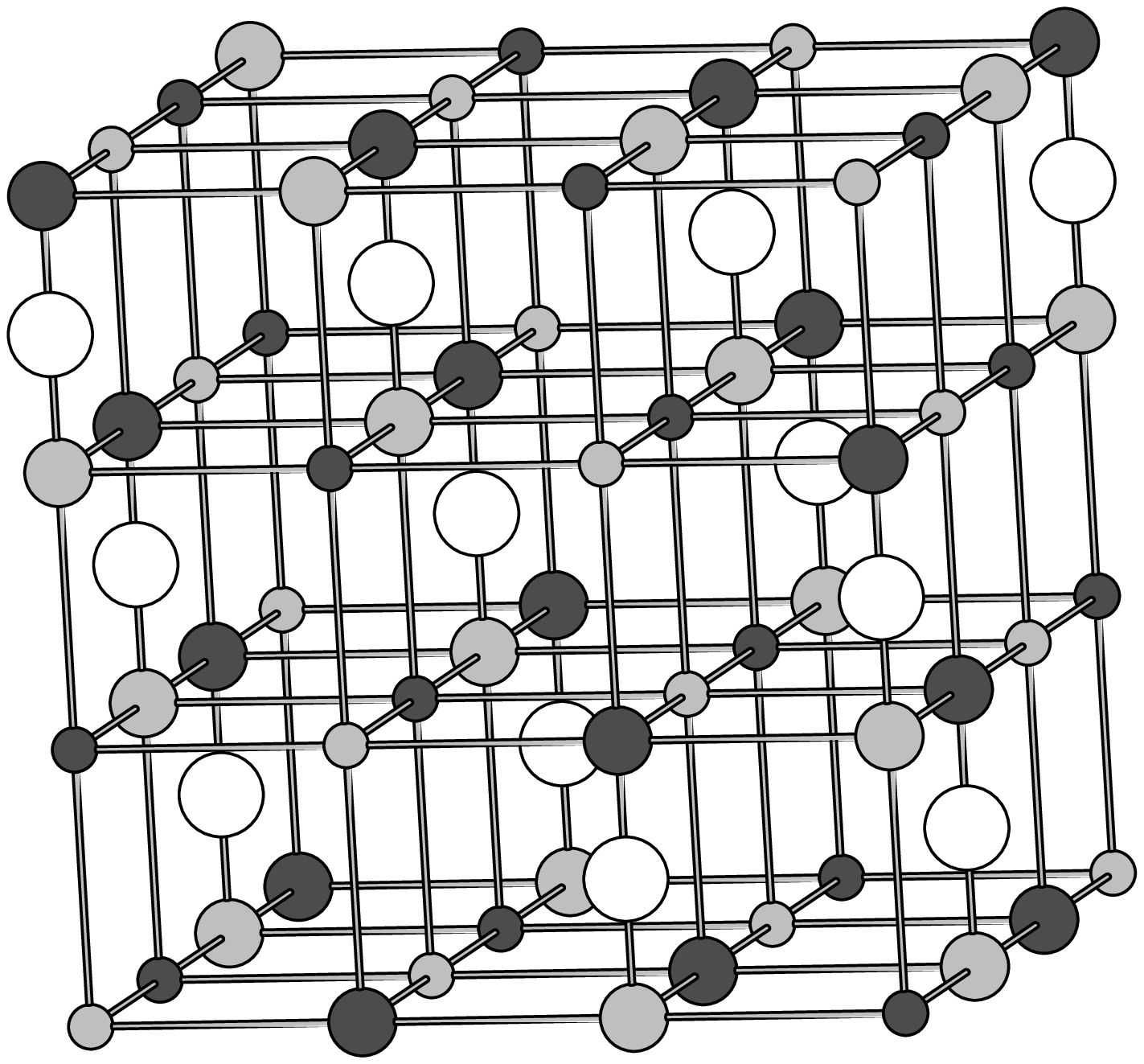}
\includegraphics*[width=0.215\textwidth]{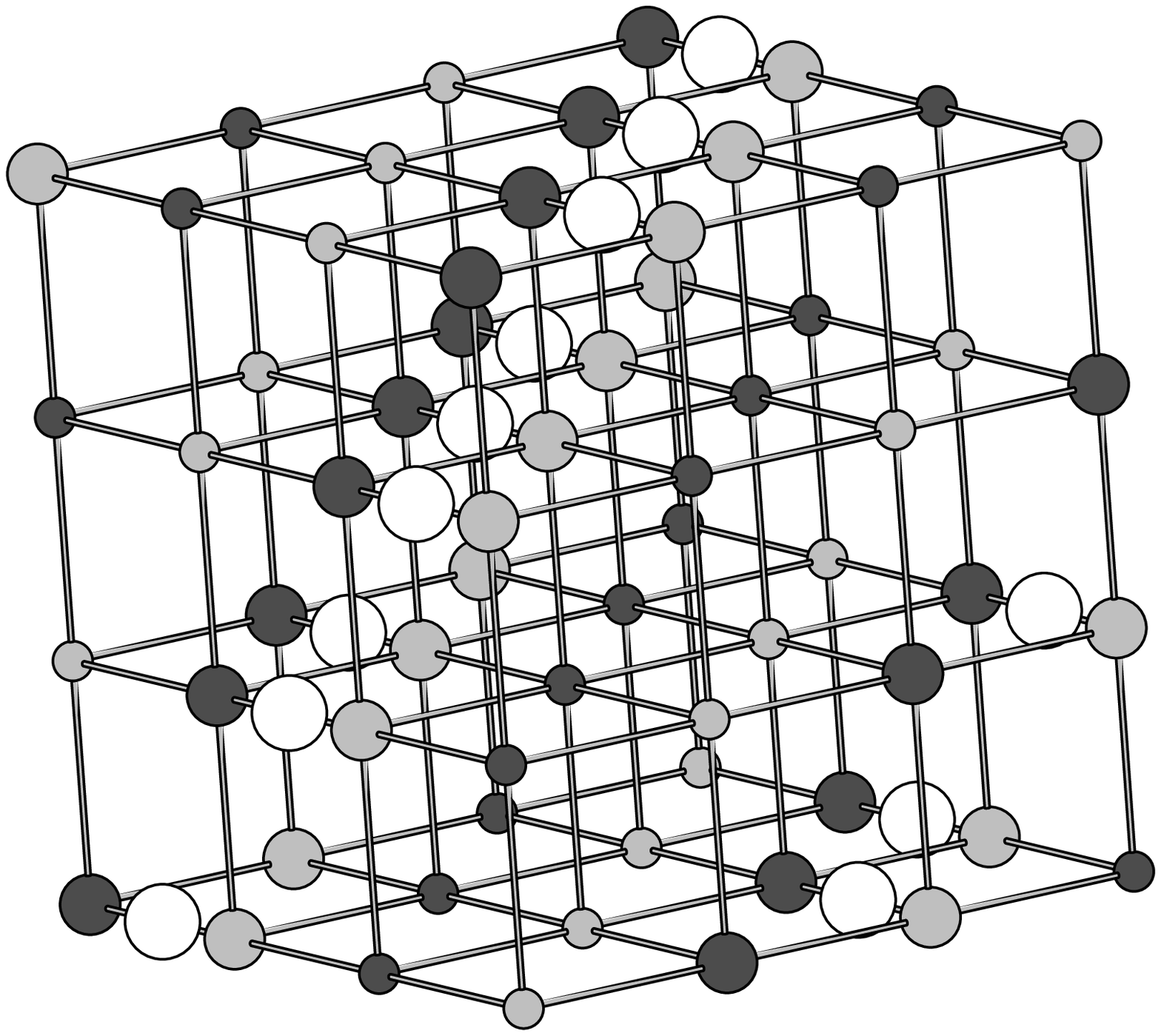}}
\caption{Schematic picture of charge ordering in BiFeO$_{2.75}$for
  vacancy configuration I. Large white spheres represent oxygen
  vacancies, Large (small) gray spheres represent ``Fe$^{2+}$''
  (``Fe$^{3+}$'') ions. Different shadings indicate the two different
  antiferromagnetic spin orientations. For simplicity all ions are
  placed at the corresponding ideal positions within the cubic
  perovskite structure. On the left side stripes of ``Fe$^{2+}$'' ions
  along the [110] direction can be identified. On the right side the
  quasi-planar arrangement of the oxygen vacancies is visible.}
\label{fig:chargeordering}
\end{figure}

The three-dimensional arrangement of ``Fe$^{2+}$'' and ``Fe$^{3+}$''
is shown schematically in Fig.~\ref{fig:chargeordering} for vacancy
configuration I. The ``Fe$^{2+}$'' cations appear on the sites
adjacent to the oxygen vacancy. The quasi-planar arrangement of oxygen
vacancies shown in Figure~\ref{fig:chargeordering} is a result of the
particular restricted supercell geometry. A more isotropic
distribution of oxygen vacancies and a smaller (and probably more
realistic) vacancy concentration would require the use of larger
supercells.

To validate our results also for a slightly smaller vacancy concentration we
repeated our calculations for a tripled unit cell containing six formula units
of BiFeO$_{3-\delta}$, corresponding to a vacancy concentration of 5.6~\% or
$\delta \approx 0.17$. In this case we describe the rhombohedral lattice using
hexagonal lattice vectors where one unit cell of the hexagonal lattice
contains three lattice points of the rhombohedral lattice. The resulting
hexagonal unit cell corresponds to space group $P3$ and in this case there are
6 groups of inequivalent oxygen anions. To reduce the computational effort, we
only treat one possible arrangement of oxygen vacancies for the tripled unit
cell which is obtained by arbitrarily removing one of the oxygen anions. This
calculation is done only for the rhombohedral bulk lattice parameters and
using the LSDA+U method.

The calculated polarization for the tripled unit cell with oxygen
vacancy is shown in Table~\ref{tab:O-vac}. Also in this case the
polarization is not significantly affected by the presence of the
oxygen vacancies. Similar to the doubled unit cells, we obtain two
distinct classes of Fe cations that can be interpreted as Fe$^{2+}$
and Fe$^{3+}$, with local Fe $d$ densities of states very similar to
those shown in Figure~\ref{fig:dos}. In this case the ratio of
``Fe$^{2+}$'' to ``Fe$^{3+}$'' is 1:2, as required by the charge
neutrality of the system. Again, the ``Fe$^{2+}$'' cations appear on
the sites adjacent to the oxygen vacancy. Therefore, although our
supercells enforce a rather artificial ordered arrangement of oxygen
vacancies and the corresponding vacancy concentrations are relatively
high, we conclude that the incorporation of oxygen vacancies in
BiFeO$_3$ leads to the formation of Fe$^{2+}$ on the sites adjacent to
the vacancy.

We now turn our attention to the effect of vacancies on the
macroscopic magnetization which is also shown in
Table~\ref{tab:O-vac}. Since the pairs of Fe$^{2+}$ cations are always
situated on neighboring positions of the magnetic lattice, and are
therefore antiferromagnetically aligned, no net magnetization results
from a ferrimagnetic arrangement of Fe$^{2+}$ and Fe$^{3+}$ cations as
occurs, for example, in magnetite. \cite{Shull/Wollan/Strauser:1951}
The magnetization is therefore still caused entirely by the small
canting of the mainly antiferromagnetically oriented magnetic moments
of the Fe cations, as shown in Ref.~\onlinecite{Ederer/Spaldin:2005}
({\it weak ferromagnetism}). \cite{Moriya:1960} Two observations can
be made by inspection of Table~\ref{tab:O-vac}. First, there is no
significant difference between the strained and unstrained
systems. Second, in some cases the magnetization is enhanced compared
to the bulk value, whereas in other cases it is decreased. The
differences are slightly more pronounced for the data set
corresponding to $U$ = 2~eV, $J$ = 0~eV than for the data set
corresponding to $U$ = 3~eV, $J$ = 1~eV, but no clear trends can be
identified. The observed changes in magnetization for the various
vacancy configurations cannot be understood by considering simple
changes in coordination between the magnetically coupled cations
(which are the same for all vacancy configurations) but rather depend
on the details of the structural relaxation of both cations and
anions. This is not surprising since the Dzyaloshinskii-Moriya
interaction which is responsible for the canting of the magnetic
moments \cite{Moriya:1960} is closely related to the superexchange
interaction which in turn is known to be very sensitive to small
structural changes. \cite{Anderson:1963} The observed changes in
magnetization caused by the presence of oxygen vacancies cannot
explain the strong increase in magnetization reported for the thin
films of BiFeO$_3$. \cite{Wang_et_al:2003}

\section{Summary and conclusions}

In summary, our calculations show that the ferroelectric polarization
in multiferroic BiFeO$_3$ is extremely insensitive to both strain and
the presence of oxygen vacancies. This insensitivity of the
polarization is due to a corresponding insensitivity of the ionic
displacements and is in striking contrast to what has been found for
most conventional perovskite ferroelectrics.  At present it is not
clear if this stability is due to a general high stability of the
ferroelectric state in BiFeO$_3$, reflected also in its large ionic
displacements and high Curie temperature ($T_C$ = 1123~K),\cite{LB:36}
if this stability is a special feature of the different mechanism
driving the ferroelectric distortion in this class of Bi-containing
multiferroic materials, namely the stereochemically active Bi lone
electron pair, \cite{Seshadri/Hill:2001} or if it is due to the
special geometry of the $R3c$ space group containing oxygen octahedra
rotations in addition to the polar displacements. Future studies of
the strain dependence of the ferroelectric polarization in the high
$T_C$ ferroelectric LiNbO$_3$ and the lone-pair active multiferroic
BiMnO$_3$ will help to solve this issue.

The incorporation of oxygen vacancies in BiFeO$_3$ leads to the
formation of Fe$^{2+}$ which can be identified by the clear
qualitative differences in the local densities of states, but the
actual charge disproportionation is small. The presence of oxygen
vacancies can affect the value of the macroscopic magnetization,
although the observed changes are too small to explain the strong
increase of the magnetization reported for thin BiFeO$_3$
film. \cite{Wang_et_al:2003} These effects of oxygen vacancies are
independent of the strain state of the system.

Recently, measurements of the electric polarization for (001), (101),
and (111) oriented films of BiFeO$_3$ (200~nm thickness) were reported
\cite{Li_et_al:2004} and the experimental data could be explained as
resulting from different projections of the same polarization
vector. This supports the notion that both magnitude and orientation
of the polarization in BiFeO$_3$ are not significantly affected by
strain, since the strain tensor differs considerably for different
film orientations. The reported value of $\sim$ 100~$\mu$C/cm$^2$ for
the (111) oriented film \cite{Li_et_al:2004} agrees well with the
polarization values calculated in this work.

\begin{acknowledgments}
The authors thank R.~Ramesh, C.~J.~Fennie, and K.~M.~Rabe for valuable
discussions and R.~Ramesh for providing the structural data for the
(001) oriented films. This work was supported by the MRSEC program of
the National Science Foundation under Award No. DMR00-80034.
\end{acknowledgments}

\appendix

\section{Equivalence of the two LSDA+U approaches for $J$ = 0}
\label{app:ldau}

In the LSDA+U approach of Dudarev {\it et al.}
\cite{Dudarev_et_al:1998} the total energy of the system is expressed
as follows (Equation (5) of Ref.~\onlinecite{Dudarev_et_al:1998},
generalized for noncollinear spin systems):
\begin{equation}
\label{dudarev}
E = E_\text{LSDA} + \frac{U_\text{eff}}{2} \left( n - \sum_{m,m',s,s'}
n^{ss'}_{mm'} n^{s's}_{m'm} \right) \quad .
\end{equation}
Here, $n^{ss'}_{mm'}$ are the elements of the orbital density matrix
and $n = \sum_{s,m} n^{ss}_{mm}$ is the total number of $d$ electrons
at the corresponding ion. Summation over all sites containing $d$
electrons has been suppressed for simplicity.

The corresponding expression for the LSDA+U approach of Anisimov,
Liechtenstein and coworkers is:
\cite{Anisimov/Aryatesiawan/Liechtenstein:1997}
\begin{multline}
\label{anisimov}
E = E_\text{LSDA} \\ + \frac{1}{2} \sum_{\{m,s\}} \left\{ \, \langle
m_1 m_3 | V_\text{ee} | m_2 m_4 \rangle \, n^{s_1 s_1}_{m_1 m_2}
n^{s_2 s_2}_{m_3 m_4} \right. \\ \left. - \, \langle m_1 m_3 |
V_\text{ee} | m_4 m_3 \rangle \, n^{s_1 s_2}_{m_1 m_2} n^{s_2
s_1}_{m_3 m_4} \right\} \\ - \frac{U}{2} n ( n - 1 ) + \frac{J}{2}
\sum_{s} n^{s} ( n^{s} - 1 ) \quad .
\end{multline}
Here, $n^{s} = \sum_{m} n^{ss}_{mm}$ is the total number of $d$
electrons with spin $s$ and $\langle m_1 m_3 | V_\text{ee} | m_2 m_4
\rangle$ are the matrix elements of the screened electron-electron
interaction. These matrix elements are defined in terms of
Slater-integrals $F^k$ as:
\begin{equation}
\label{vee}
\langle m_1 m_3 | V_\text{ee} | m_2 m_4 \rangle = \sum_k a_k(m_1,m_3,m_2,m_4)
\, F^k \quad ,
\end{equation}
with
\begin{multline}
\label{angular}
a_k(m_1,m_3,m_2,m_4) \\ = \frac{4 \pi}{2k+1} \sum_{q = -k}^{k} \langle l
m_1 | Y_{kq} | l m_2 \rangle \langle l m_3 | Y_{kq}^* | l m_4 \rangle
\quad .
\end{multline}
The Slater integrals are related to the parameters $U$ and $J$ via $U
= F^0$, $J = (F^2 + F^4)/14$, where one usually sets $F^2/F^4 =
0.625$. For $J = 0$ the only nonvanishing term in the sum of
Eq.~(\ref{vee}) is for $k=0$, and one can see from Eq.~(\ref{angular})
that $a_0(m_1,m_2,m_3,m_4) = \delta_{m_1 m_2} \delta_{m_3 m_4}$. For
$J=0$ the matrix elements (\ref{vee}) are therefore simply given by:
\begin{equation}
\label{j=0}
\langle m_1 m_3 | V_\text{ee} | m_2 m_4 \rangle  = U \delta_{m_1 m_2}
\delta_{m_3 m_4} \quad .
\end{equation}
Using (\ref{j=0}) in (\ref{anisimov}) one obtains:
\begin{equation}
E = E_\text{LSDA} + \frac{U}{2} \left( n - \sum_{m,m',s,s'}
n^{ss'}_{mm'} n^{s's}_{m'm} \right) \quad ,
\end{equation}
which is just Eq.~(\ref{dudarev}) with $U_\text{eff} = U$. The LSDA+U
approach of Dudarev {\it et al.} \cite{Dudarev_et_al:1998} is
therefore included in the approach of
Ref.~\onlinecite{Anisimov/Aryatesiawan/Liechtenstein:1997} as the
special case $J=0$.


\bibliography{/home/ederer/main/tex/literature}

\end{document}